\documentclass[12pt]{article}
\usepackage{latexsym,amssymb}
\usepackage{graphicx}

\makeatletter
\newfont{\footsc}{cmcsc10 at 8truept}
\newfont{\footbf}{cmbx10 at 8truept}
\newfont{\footrm}{cmr10 at 10truept}
\renewcommand{\ps@plain}{%
\renewcommand{\@oddfoot}{\footsc the electronic journal of combinatorics
  {\footbf } , \#R\hfil\footrm\thepage}}
\makeatother
\usepackage{amsmath,amssymb,amsfonts,latexsym,theorem}

\numberwithin{equation}{section}
\newtheorem{theorem}{Theorem}[section]
\newtheorem{proposition}[theorem]{Proposition}
\newtheorem{corollary}[theorem]{Corollary}
\newtheorem{conjecture}[theorem]{Conjecture}

{\theorembodyfont\rmfamily
\newtheorem{definition}[theorem]{Definition}

}

\def\mn{\text{-}}

\def\qed{\hfill $\Box$}

\newenvironment{proof}{\noindent{\scshape Proof.}}{\qed}

\begin{document}

\def\sof{\hfill\rule{2mm}{2mm}}
\def\SS{\mathcal S}
\def\qq{{\bold q}}
\def\mn{\text{-}}

\title{\sc A matrix representation of graphs and its spectrum as a graph
invariant}

\author{\textbf{David Emms}\\
Department of Computer Science\\
University of York, York YO10 5DD, U.K.\\
\texttt{demms@cs.york.ac.uk} \and
\textbf{Edwin R. Hancock}\\
Department of Computer Science\\
University of York, York YO10 5DD, U.K.\\
\texttt{erh@cs.york.ac.uk}
\and
\textbf{Simone Severini} \\
Department of Mathematics and Department of Computer Science\\
University of York, York YO10 5DD, U.K.\\
\texttt{ss54@york.ac.uk}
\and
\textbf{Richard C. Wilson} \\
Department of Computer Science\\
University of York, York YO10 5DD, U.K.\\
\texttt{richard.wilson@cs.york.ac.uk} }


\date{\small Submitted: ;  Accepted: .\\
\small 2000 Mathematics Subject Classification: 05E30, 05C60}

\maketitle

\begin{abstract}
We use the line digraph construction to associate an orthogonal
matrix with each graph. From this orthogonal matrix, we derive two
further matrices. The spectrum of each of these three matrices is
considered as a graph invariant. For the first two cases, we compute
the spectrum explicitly and show that it is determined by the
spectrum of the adjacency matrix of the original graph. We then show
by computation that the isomorphism classes of many known families
of strongly regular graphs (up to 64 vertices) are characterized by
the spectrum of this matrix. We conjecture that this is always the
case for strongly regular graphs and we show that the conjecture is
not valid for general graphs. We verify that the smallest regular
graphs which are not distinguished with our method are on 14
vertices.
\end{abstract}

\section{Introduction}

Graphs are often conveniently represented using matrices, for
example, the adjacency matrix, the Laplacian matrix, \emph{etc.}
\cite{g}. Many important properties of a graph are encoded in the
eigenvalues of the matrix representation. However, eigenvalues
generally fail to separate isomorphism classes. In this paper, we
consider some matrix representations inspired by the notion of
coined quantum walks \cite{a}. Since strongly regular graphs give
rise to relatively large sets of non-isomorphic graphs which are
cospectral with respect to the commonly used matrix representations,
we use these graphs as a testing ground for the representations that
we define. Let $G$ be a graph and let $U$ be the orthogonal matrix
inducing a coined
quantum walk on $G$, where the coins are Grover matrices. For every given $k$%
, we define a digraph, $D$, with an arc $(i,j)$ if and only if $%
U_{i,j}^{k}>0 $. For $k=1$, we express the adjacency eigenvalues of
$D$ in terms of the ones of $G$. When considering strongly regular
graphs, we describe how to construct directly $D$ from $G$. For
strongly regular graphs, we conjecture that the eigenvalues of the
adjacency matrix of $D$ distinguish $G$ from its cospectral mates
for $k=3$. We verify this conjecture numerically for all strongly
regular graphs up to $64$ vertices contained in the tables of Spence
\cite{s} (the graphs srg$(16,9,4,6)$ were obtained from the tables
of McKay \cite{mc}). By providing counterexamples, we show that our
method fails to distinguish general graphs. Hopefully these
counterexamples will help understanding what graphs are cospectral
with respect to the representations described here.

\bigskip

The remainder of the paper is organized as follows. In Section 2, we
recall the definition of the orthogonal matrix $U$ and give a
formula for its eigenvalues. We define digraphs derived from powers
of $U$, and propose using these digraphs to distinguish the original
graph from its cospectral mates. In Section 3, we focus on strongly
regular graphs. We describe some structural properties of the
digraph $D$ obtained from $U^{3}$. In Section
4, we list the sets of strongly regular graphs for which the eigenvalues of $%
D$ successfully distinguish a graph from its cospectral mates. We
conclude with counterexamples involving general graphs.

\bigskip

Recently, dynamical processes based on quantum evolution have been
considered for attacking the graph isomorphism problem \cite{gu,
sjc}. In this context, Shiau \emph{et al.} \cite{sjc} gave evidence
that quantum walks alone are not successful. The setting they
describe is as follows. Let
$\{|j\rangle :1\leq j\leq n\}$ be an orthonormal basis of a Hilbert space ($%
|j\rangle $ is a ray and $\langle i|$ is the linear functional which
maps each $\left\vert j\right\rangle $ to the usual inner product,
which is
denoted by $\langle i%
{\vert}%
j\rangle $). Given a graph $G$, let us define the Hamiltonian
$H=-\sum M(G)_{a,b}c_{a}^{\dagger }c_{b}$, where the operator
$c_{a}^{\dagger }c_{b}$ is given by $\langle i|c_{a}^{\dagger
}c_{b}|j\rangle =\delta _{i,a}\delta
_{b,j}$ and $M(G)$ is the adjacency matrix of $G$. The evolution for a time $%
t$ of the initial states $\{|\psi _{j}(0)\rangle =|j\rangle :1\leq
j\leq n\}$
is governed by the Schr\"{o}dinger equation $i\frac{d|\psi _{j}(0)\rangle }{%
dt}=H|\psi _{j}(0)\rangle $. In \cite{sjc}, it was shown that the
matrix whose entries are $O_{i,j}=\langle \psi _{i}(0)|\psi
_{j}(t)\rangle $ do not help in distinguishing pairs of strongly
regular graphs with the same set of
parameters. However, by making use of some techniques described in \cite{ru}%
, it was also given an alternative method which distinguished
numerically all known strongly regular graphs up to $29$ vertices.
It would be interesting to investigate potential connections between
\cite{sjc} and the present paper.

\section{Representing graphs with orthogonal matrices}

Let $G=(V,E)$ be a simple undirected graph (that is $G$ is loopless
and without multiple edges). Let $D_{G}=(V,A)$ be the directed graph
obtained
from $G$ by replacing every edge $\{i,j\}\in E(G)$ with the pair of arcs $%
(i,j)$ and $(j,i)$. The degree of a vertex $i$ is denoted by $d(i)$.
A graph is said to be $k$-regular if every of its vertices has
degree $k$. Unless otherwise stated, in this paper we will consider
only graphs with minimum degree $3$.

\begin{definition}
Given a graph $G$, we denote by $U(G)$\ the matrix defined as follows:%
\[
\text{for all }(i,j),(k,l)\in A(D_{G}),
\]%
\[
U(G)_{(i,j),(k,l)}:=\left\{
\begin{tabular}{ll}
$\frac{2}{d(j)}-\delta_{i,l},$ & if $j=k;$ \\
$0,$ & otherwise$.$%
\end{tabular}
\right.
\]
\emph{\ }
\end{definition}

The matrix $U(G)$ induces a coined quantum walk on $G$, where the
coins are Grover matrices \cite{a}. Here are some basic observations
about $U(G)$:

\begin{itemize}
\item If $|E(G)|=m$ then $U(G)$ is $2m\times2m$ and the vertices of degree
zero do not contribute to the dimension of $U_{G}$.

\item Given $i\in V(G)$, if $d(i)=1$ and $\{i,j\}\in E(G)$, then $%
U(G)_{(j,i),(i,j)}=1$, and this is the unique non-zero entry in the
row indexed by $(j,i)$ and in the column indexed by $(i,j)$.

\item Given $j\in V(G)$, if $d(j)=2$ and $\{i,j\},\{j,k\}\in E(G)$, then $%
U(G)_{(i,j),(j,i)}=U(G)_{(k,j),(j,k)}=0$ and $%
U(G)_{(i,j),(j,k)}=U(G)_{(k,j),(j,i)}=1$. So, if $G$ is $2$-regular,
the matrix $U(G)$ has exactly a one in every row and every column
and it is then a permutation matrix.

\item Since $G$ is undirected, $U(G)_{(i,j),(j,i)}\neq0$ and $%
U(G)_{(j,i),(i,j)}\neq0$ for every $\{i,j\}\in E(G)$. However, we
have assumed that the minimum degree of $G$ is $3$, and then $U(G)$
is not
symmetric. In fact, for three distinct vertices $i,j,l\in V(G)$ such that $%
\{i,j\},\{j,l\}\in E(G)$, we have $U(G)_{(i,j),(j,l)}\neq0$, but $%
U(G)_{(j,l),(i,j)}=0$ since $i\neq l$.
\end{itemize}

The \emph{support} of an $n\times n$ matrix $M$, denoted by
$\underline{M}$, is the $n\times n$ $\left( 0,1\right) $-matrix with
elements defined as
follows:%
\[
\underline{M}_{i,j}:=\left\{
\begin{tabular}{ll}
$1$ & if $M_{i,j}\neq0;$ \\
$0$ & otherwise.%
\end{tabular}
\right.
\]

Let $M(G)$ be the adjacency matrix of a graph (digraph) $G$. The
\emph{line digraph} of a digraph $D$, denoted by
$\overrightarrow{L}D$, is the digraph
defined as follows: $V(\overrightarrow{L}D)=A(D)$ and $((i,j),(k,l))\in A(%
\overrightarrow{L}D)$ if and only if $j=k$. If $d(i)>2$ for every
$i\in V(G)$
then one can verify that $\underline{U(G)}=M(\overrightarrow{L}D_{G})$ \cite%
{si}.

\bigskip

Graphs (or digraphs) $G$ and $H$ are \emph{isomorphic} (written
$G\cong H$)
if there is a permutation matrix $P$ such that $M(G)=PM(H)P^{-1}$, where $%
M(G)$ and $M(H)$ are the adjacency matrices of the graphs $G$ and
$H$, respectively. Harary and Norman \cite{hn} proved that if $D$
and $F$ are digraphs without sources or sinks then
$\overrightarrow{L}D\cong
\overrightarrow{L}F$ if and only if $D\cong F$ ($i\in V(D)$ is a \emph{source%
} if there is no $j$ such that $(j,i)\in A(D)$; a \emph{sink }if
there is no $j$ such that $(i,j)\notin A(D)$). Since $D_{G}$ is
without sources or
sinks, it follows that $G\cong H$ if and only if $\underline{U(G)}=Q%
\underline{U(H)}Q^{-1}$, for some permutation matrix $Q$. According
to Definition 1, this fact is sufficient to ensure that $G\cong H$
if and only if $U(G)=QU(H)Q^{-1}$, for some permutation matrix $Q$.
In words, $U(G)$ \textquotedblleft faithfully\textquotedblright
\emph{\ }represents $G$.

\bigskip

The \emph{spectrum} of a matrix $M$ is the multiset of the eigenvalues of $M$%
, and is denoted by Sp$(M)=\{[\lambda_{i}]^{m_{i}}:1\leq i\leq m\}$, where $%
\lambda_{1},\lambda_{2},...,\lambda_{k}$ are the eigenvalues of $M$ and $%
m_{1},m_{2},...,m_{k}$ their respective multiplicities. Let $T(G)$
be the
matrix with $ij$-th entry%
\[
T(G)_{i,j}:=\left\{
\begin{tabular}{ll}
$\frac{1}{d(j)},$ & if $M(G)_{i,j}=1;$ \\
$0,$ & otherwise.%
\end{tabular}
\ \ \right.
\]
We can obtain Sp$(U(G))$ from Sp$(T(G))$, as shown in the next
result.

\begin{proposition}
\label{ft}Let $G$ be a graph on $n$ vertices and $m$ edges. The
matrix $U(G)$ has $2n$ eigenvalues of the form
\[
\lambda=\lambda_{T}\pm i\sqrt{1-\lambda_{T}^{2}},
\]
where $\lambda_{T}$ is an eigenvalue of the matrix $T(G)$. The remaining $%
2(m-n)$ eigenvalues of $U(G)$ are $\pm1$ with equal multiplicities.
\end{proposition}

\begin{proof}
Let $M=M(G)$, $U=(G)$ and $T=T(G)$. Since $G$ is undirected, we have $%
M_{i,j}^{2}=M_{i,j}M_{j,i}=M_{i,j}$. By Definition 1,%
\[
U_{(i,j),(k,l)}=M_{i,j}M_{k,l}\delta_{j,k}\left(
\frac{2}{d(k)}-\delta _{i,l}\right) .
\]
Let $\mathbf{u}$ be an eigenvector of $U$ with eigenvalue $\lambda$
(notice that the entries of $\mathbf{u}$ have two indices as it is
for the rows and the columns of $U$). By making use of this
equation, we can write
\begin{align*}
\lambda u_{\left( i,j\right) } & =\sum_{k,l}U_{(i,j),(k,l)}u_{\left(
k,l\right) } \\
& =2\frac{M_{i,j}}{d(j)}\sum_{l}M_{j,l}u_{\left( j,l\right)
}-M_{i,j}u_{\left( j,i\right) }.
\end{align*}
Let $\mathbf{t}$ be an eigenvector of $T$ with eigenvalue
$\lambda_{T}$. Define
\[
u_{\left( i,j\right) }=\frac{M_{i,j}t_{j}}{d(j)}-\lambda^{\ast}\frac {%
M_{i,j}t_{i}}{d(i)}.
\]
Then
\begin{align*}
& 2\frac{M_{i,j}}{d(j)}\sum_{l}M_{j,l}u_{\left( j,l\right)
}-M_{i,j}u_{\left( j,i\right) } \\
& =\frac{M_{i,j}t_{j}}{d(j)}(2\lambda_{T}-\lambda^{\ast})-M_{i,j}\frac{t_{i}%
}{d(i)}.
\end{align*}
Since $U$ is real-orthogonal, its eigenvalues have magnitude $1$. It
follows that
\[
\lambda u_{\left( i,j\right) }=\lambda\left( \frac{M_{i,j}t_{j}}{d_{j}}%
(2\lambda^{\ast}\lambda_{T}-(\lambda^{\ast})^{2})-\lambda^{\ast}M_{i,j}\frac{%
t_{i}}{d_{i}}\right) .
\]
Therefore, $\mathbf{u}$ is an eigenvector of $U$ if $2\lambda^{\ast}%
\lambda_{T}-(\lambda^{\ast})^{2}=1$ or
$\lambda+\lambda^{\ast}=2\lambda_{T}$ ($\lambda^{\ast}$ is the
complex conjugate of $\lambda$). It is easy to verify that the
remaining eigenvalues are $\pm1$.
\end{proof}

\bigskip

Two graphs $G$ and $H$ are said to be \emph{cospectral} (or \emph{isospectral%
}), with respect to a given matrix representation, if the spectra of
the matrices representing the graphs are identical (see,
\emph{e.g.}, \cite{w}).
For example, let $G$ and $H$ such that%
\[
M(G)=\left(
\begin{array}{ccccc}
0 & 1 & 1 & 0 & 0 \\
1 & 0 & 0 & 1 & 0 \\
1 & 0 & 0 & 1 & 0 \\
0 & 1 & 1 & 0 & 0 \\
0 & 0 & 0 & 0 & 0%
\end{array}
\right)
\]
and%
\[
M(H)=\left(
\begin{array}{ccccc}
0 & 0 & 0 & 0 & 1 \\
0 & 0 & 0 & 0 & 1 \\
0 & 0 & 0 & 0 & 1 \\
0 & 0 & 0 & 0 & 1 \\
1 & 1 & 1 & 1 & 0%
\end{array}
\right) .
\]
Then Sp$(M(G))=$ Sp$(M(H))$. The following corollary is a
straightforward consequence of Proposition \ref{ft}.

\begin{corollary}
Given graphs $G$ and $H$, \emph{Sp}$(U(G))=$ \emph{Sp}$(U(H))$ iff \emph{Sp}$%
(T(G))=$ \emph{Sp}$(T(H))$.
\end{corollary}

If $G\cong H$ then Sp$(M(G))=$ Sp$(M(H))$, but the converse is not
necessarily true. This is the case for the graphs $G$ and $H$
considered in the above example, since Sp$(M(G))=$
Sp$(M(H))=\{[-2]^{1},[0]^{3},[2]^{2}\}$ but, of course, $G\ncong H$.
Also, Sp$(T(G))=$ Sp$(T(H))$ does not imply
that $G\cong H$. For instance, if $G$ is a $k$-regular graph then Sp$%
(T(G))=\{[\frac{1}{k}\lambda _{i}]^{m_{i}}:1\leq i\leq k\}$, where Sp$%
(M(G))=\{[\lambda _{i}]^{m_{i}}:1\leq i\leq k\}$. So, for two
$k$-regular
graphs $G$ and $H$, Sp$(M(G))=$ Sp$(M(H))$ if and only if Sp$(T(G))=$ Sp$%
(T(H))$, or, equivalently, Sp$(U(G))=$ Sp$(U(H))$. Two non-isomorphic $k$%
-regular cospectral graphs $G$ and $H$ are then not distinguished by
the spectra of $U(G)$ and $U(H)$.

\bigskip

Now, let $M$ and $N$ be matrices such that Sp$(M)=$
Sp$(N)=\{[\lambda
_{i}]^{m_{i}}:1\leq i\leq n\}$. For any natural number $p$, Sp$(M^{p})=$ Sp$%
(N^{p})=\{[\lambda _{i}^{p}]^{m_{i}}:1\leq i\leq n\}$, but it is not
necessarily the case that Sp$(\underline{M^{p}})=$
Sp$(\underline{N^{p}})$. This is again the case of the graphs $G$
and $H$ in the above example:
\[
\underline{M(G)^{2}}=\left(
\begin{array}{ccccc}
1 & 0 & 0 & 1 & 0 \\
0 & 1 & 1 & 0 & 0 \\
0 & 1 & 1 & 0 & 0 \\
1 & 0 & 0 & 1 & 0 \\
0 & 0 & 0 & 0 & 0%
\end{array}%
\right) ;
\]%
\[
\underline{M(H)^{2}}=\left(
\begin{array}{ccccc}
1 & 1 & 1 & 1 & 0 \\
1 & 1 & 1 & 1 & 0 \\
1 & 1 & 1 & 1 & 0 \\
1 & 1 & 1 & 1 & 0 \\
0 & 0 & 0 & 0 & 1%
\end{array}%
\right) .
\]%
Sp$(\underline{M(G)^{2}})=\{[0]^{3},[2]^{2}\}$ and Sp$(\underline{M(H)^{2}}%
)=\{[0]^{3},[1],[4]\}$. The entry $M(G)_{i,j}^{p}$ equals the number
of
paths of length $p$ between vertices $i$ and $j$. Note that $%
M(G)_{1,1}^{2}=2 $ and $\underline{M(G)^{2}}_{1,1}=1$. This implies
that matrix $\underline{M(G)^{p}}$ contains generally less
information about the structure of a graph $G$, since
$\underline{M(G)^{p}}_{i,j}=1$ if and only if the number of paths of
length $p$ between vertices $i$ and $j$ is non-zero. Despite this
fact, it is legitimate to ask the following question: given graphs
$G$ and $H $ such that Sp$(M(G))=$ Sp$(M(H))$ what conditions imply
that Sp$(\underline{M(G)^{p}})\neq $ Sp$(\underline{M(H)^{p}})$, for
some $p$? If a graph $G$ with diameter $p$ is connected and
non-bipartite then $\underline{M(G)^{p}}=J $, where $J$ denotes the
all-ones matrix. As a
consequence, if two graph $G$ and $H$ have diameter $p$ then Sp$(\underline{%
M(G)^{q}})=$ Sp$(\underline{M(H)^{q}})$ for every $q\geq p$, even if Sp$%
(M(G))\neq $ Sp$(M(H))$. The spectra Sp$(\underline{M(G)^{q}})$ and Sp$(%
\underline{M(H)^{q}})$ are indeed of no use in distinguishing $G$
and $H$ if these graphs have diameter $2$.

\bigskip

Given a graph $G$, since the matrix $U(G)$ is real-orthogonal, the
support
of $U(G)^{p}$, for some $p$, is presumably different from the support of $M(%
\overrightarrow{L}D_{G})^{p}$. This is because of the contribution
of the negative entries in $U(G)$. On the basis of this observation,
we define the following matrix.

\begin{definition}
Given a graph $G$, we denote by $S^{+}(U(G)^{p})$\ the matrix
defined as
follows:\emph{\ }%
\[
\begin{tabular}{l}
$S^{+}(U(G)_{i,j}^{p}):=\left\{
\begin{tabular}{ll}
$1$ & if $U(G)_{i,j}^{p}>0;$ \\
$0$ & otherwise$.$%
\end{tabular}
\ \right. $%
\end{tabular}
\]
\end{definition}

Suppose that Sp$(M(G))=$ Sp$(M(H))$ and $G\ncong H$. What conditions
need $G$
and $H$ to satisfy in order that Sp$(S^{+}(U(G)^{p}))\neq $ Sp$%
(S^{+}(U(H)^{p}))$ for some $p\geq 2$? Strongly regular graphs seem
to provide a good testing ground to approach this question. Strongly
regular graphs have been investigated in many different contexts,
including group theory, algebraic graph theory, design of
experiments, finite geometries, error-correcting codes, \emph{etc.}
(see, \emph{e.g.}, \cite{br, c}). The fastest known algorithm for
testing the isomorphism of strongly regular
graphs on $n$ vertices was designed by Spielman \cite{sp} and runs in $%
n^{O(n^{1/3}\log n)}$ number of steps. A \emph{strongly regular
graph} with parameters $(n,d,r,s)$ (for short, a srg$(n,d,r,s)$) is
a $d$-regular graph on $n$ vertices such that any two adjacent
vertices have exactly $r$ common neighbours and any two nonadjacent
vertices have exactly $s$ common neighbours \cite{c}. Strongly
regular graphs are interesting in our context principally for the
following two reasons:

\begin{itemize}
\item For the strongly regular graphs $G$ and $H$ with identical sets of
parameters, we have Sp$(M(G))=$ Sp$(M(H))$ even if $G\ncong H$. In
particular, the adjacency matrix of a srg$(n,d,r,s)$ has exactly
three eigenvalues, $[d]^{1}>[e^{+}]^{m^{+}}\geq \lbrack
e^{-}]^{m^{-}}$, such that
$e^{+}=\frac{1}{2}(r-s+\sqrt{\Delta })$ and $e^{-}=\frac{1}{2}(r-s-\sqrt{%
\Delta })$, with $\{m^{+},m^{-}\}=\frac{1}{2}(n-1\pm \frac{2k+(n-1)(s-r)}{%
\sqrt{\Delta }})$ and $\Delta =\left( s-r\right) ^{2}+4\left(
d-s\right) $.

\item The diameter of a connected strongly regular graph is $2$ (see, \emph{%
e.g.}, \cite{c}).
\end{itemize}

\section{On the spectra of $S^{+}(U(G))$, $S^{+}(U(G)^{2})$ and $%
S^{+}(U(G)^{3})$}

In this section, we give formulas to compute the eigenvalues of
$S^{+}(U(G))$ and $S^{+}(U(G)^{2})$ from the eigenvalues of $M(G)$,
when $G$ is strongly regular. It follows that neither $S^{+}(U(G))$
nor $S^{+}(U(G)^{2})$ are of any help in distinguishing $G$. The
eigenvalues of $S^{+}(U(G)^{3})$ do not depend entirely on the
eigenvalues of $M(G)$ and it seems difficult to give a formula for
Sp$(S^{+}(U(G)^{3}))$ in terms of properties of $G$. On the other
hand we show how to construct $S^{+}(U(G)^{3})$ from $G$. We
conjecture that Sp$(S^{+}(U(G)^{3}))$ distinguishes $G$ from its
cospectral mates.

\bigskip

Given a digraph $D$, let $P(D,x)=\det (xI-M\left( D\right) )$ be the
characteristic polynomial of $M\left( D\right) $. It is known that
\cite{lz,
r}%
\[
P(\overrightarrow{L}D,x)=x^{%
{\vert}%
A\left( D\right)
{\vert}%
-%
{\vert}%
V\left( D\right)
{\vert}%
}P\left( D,x\right) .
\]%
Then
\[
\text{Sp}(\underline{U(G)})=\text{ Sp}(M(G))\cup
\{[0]^{nk-(1+m^{+}+m^{-})}\}.
\]

The following two propositions show that the spectra of $S^{+}(U(G))$ and $%
S^{+}(U(G)^{2})$ are determined by the parameters of $G$. Therefore
the spectra of these matrices do not characterize $G$ itself.

\begin{proposition}
Let $G$ be a $k$-regular graph on $n$ vertices. Let $\mathbf{e}$ be
an eigenvector of $M(G)$ with eigenvalue $\lambda_{M}$ and let
$\mathbf{u}$ be
an eigenvector of $S^{+}(U(G))$ with eigenvalue $\lambda$. Then $S^{+}(U(G))$ has $%
2n$ eigenvalues of the form
\[
\lambda=\frac{\lambda_{M}}{2}\pm i\sqrt{k-1-\lambda_{M}^{2}/4},
\]
with eigenvectors having entries
\[
u_{i,j}=M(G)_{i,j}\mathbf{e}_{j}-\frac{1}{\lambda}M(G)_{i,j}e_{i}.
\]
The remaining $n(k-2)$ eigenvalues of $S^{+}(U(G))$ take the values
$\pm1$.
\end{proposition}

\begin{proof}
Observe that
\[
S^{+}(U(G)_{(i,j),(k,l)})=(1-\delta_{i,l})M_{i,j}M_{k,l}\delta_{j,k}.
\]
We then have%
\begin{align*}
& \sum_{(x,y)}S^{+}(U(G)_{(a,b),(x,y)})u_{x,y} \\
& =M_{a,b}\sum_{y}M_{b,y}u_{b,y}-M_{a,b}u_{b,a} \\
& =M_{a,b}\sum_{y}M_{b,y}e_{y}-\frac{1}{\lambda}M_{a,b}e_{b} \\
& \sum_{y}M_{b,y}-M_{a,b}e_{a}+\frac{1}{\lambda}M_{a,b}e_{b} \\
& =M_{a,b}e_{b}(\lambda-(k-1)/\lambda)-M_{a,b}e_{a} \\
& =\lambda\left( M_{a,b}e_{b}-\frac{1}{\lambda}M_{a,b}e_{a}\right)
\end{align*}
Then these eigenvalues and eigenvectors account for $2n$ of the
eigenvalues
and eigenvectors of $S^{+}(U(G))$. It is easy to verify that the remaining $%
n(k-2)$ eigenvalues take the values $\pm1$.
\end{proof}

\begin{proposition}
Let $G$ be a $k$-regular graph on $n$ vertices. Let $\mathbf{e}$ be
an eigenvector of $M(G)$ with eigenvalue $\lambda_{M}$ and let
$\mathbf{u}$ be
an eigenvector of $S^{+}(U(G)^{2})$ with eigenvalue $\lambda$. Then $%
S^{+}(U(G)^{2})$ has $2n$ eigenvalues of the form%
\[
\lambda=\frac{\lambda_{M}^{2}}{2}+2-k\pm i\lambda_{M}\sqrt{%
k-1-\lambda_{M}^{2}/4}
\]
with eigenvectors having entries
\[
u_{i,j}=M(G)_{i,j}e_{j}-\frac{\lambda^{\ast}-2+k}{\lambda_{M}(k-1)}%
A_{i,j}e_{i}.
\]
The remaining $n(k-2)$ eigenvalues of $S^{+}(U(G)^{2})$ take the
value $2$.
\end{proposition}

\begin{proof}
Observe that
\[
S^{+}(U(G)^{2})=M_{i,j}M_{j,k}M_{k,l}(2\delta_{i,k}\delta_{j,l}+1-\delta
_{i,k}-\delta_{j,l}).
\]
We then have{\small
\begin{align*}
\lefteqn{\sum_{(x,y)}S^{+}(U(G)_{(a,b),(x,y)})u_{x,y}} \\
&
=\sum_{x,y}M_{a,b}M_{b,x}M_{x,y}(2\delta_{a,x}\delta_{b,y}+1-\delta
_{a,x}-\delta_{b,y}) \\
& \left( M_{x,y}e_{y}-\frac{\lambda_{M}(\lambda^{\ast}-2+k)}{k-1}%
M_{x,y}e_{x}\right) \\
& =M_{a,b}e_{b}(\lambda_{M}^{2}-\lambda^{\ast})-M_{a,b}e_{a} \\
& \left(
\frac{(2-k)(\lambda^{\ast}-2+k)+\lambda_{M}^{2}(k-1)}{\lambda
_{M}(k-1)}\right) \\
& =M_{a,b}e_{b}\lambda-M_{a,b}\mathbf{e}_{a} \\
& \left( \frac{(\lambda^{\ast}-2+k)(2-k+\lambda-2+k)}{\lambda_{M}(k-1)}%
\right) \\
& =\lambda u_{a,b}.
\end{align*}
}We can also write
\begin{align*}
& \sum_{(x,y)}S^{+}(U(G)_{(a,b),(x,y)})u_{(x,y)} \\
& =2M_{a,b}u_{a,b}+M_{a,b}\sum_{x,y}M_{b,x}M_{x,y}u_{x,y} \\
&
-M_{a,b}\sum_{y}M_{b,a}M_{a,y}u_{a,y}-M_{a,b}\sum_{x}M_{b,x}M_{x,b}u_{x,b}.
\end{align*}
If
\[
\begin{tabular}{lll}
$\sum_{y}M_{a,y}u_{a,y}=0$ & and & $\sum_{y}M_{x,b}u_{x,b}=0$%
\end{tabular}
\]
then
\[
\sum_{(x,y)}S^{+}(U(G)_{(a,b),(x,y)})u_{x,y}=2M_{a,b}u_{a,b}
\]
and $\mathbf{u}$ is an eigenvector with eigenvalue $2$, provided that $%
M_{a,b}=0$ and then $u_{a,b}=0$.
\end{proof}

\bigskip

The following observation allows us to construct $S^{+}(U(G)^{3})$
directly
from a strongly regular graph $G$, without the need of first constructing $%
U(G)$.

\begin{proposition}
Let $G$ be a srg$(n,k,r,s)$. Then $S^{+}(U(G)_{(i,j),(l,m)}^{3})=1$
if and only if one of the following conditions holds:

\begin{enumerate}
\item $i=m$, $j\neq l$ and
\[
s+\left( r-s\right) M(G)_{j,l}-k+\frac{k^{2}}{4}>0
\]
(which always holds if $i=m$, $j\neq l$ and $k>4$);

\item $i=l$, $m\neq j$ and $M(G)_{j,m}<\frac{2r}{k}$;

\item $i=l$ and $m=j$;

\item $i\neq l$, $m=j$ and $M(G)_{i,l}<\frac{2r}{k}$;

\item $i\neq l$, $i\neq m$, $j\neq l$, $j\neq m$ and%
\[
\frac{2}{k}\left[ s+(r-s)M(G)_{j,l}\right] >M(G)_{i,l}+M(G)_{j,m}.
\]
\end{enumerate}
\end{proposition}

\begin{proof}
Let $G$ be a srg$(n,k,r,s)$. Let $A=M(G)$ and $U=U(G)$. We define the \emph{%
amplitude} of a given path to be the product of the entries of $U$
corresponding to the transitions along the path. The entry $%
S^{+}(U(G)_{(i,j),(l,m)}^{3})$ is given by the sum of the amplitudes
of all possible paths of length three from $(i,j)$ to $(l,m)$. In
what follows we count all these paths to give conditions for
$S^{+}(U(G)_{(i,j),(l,m)}^{3})=1$. We begin by observing that $i\neq
j$ and $l\neq m$ since $G$ is a simple graph. The cases in the
following table completely describe the possible ways in which the
vertices $i,j,l,\text{ and }m$ can be identified with one
another:%
\[
\begin{tabular}{||l|l|l||}
\hline $i=m(\neq l)$ & $i=l(\neq m)$ & $i\neq m$ and $i\neq l$
\\ \hline
Case A: $j=l$ & Case C: $j=m$ & Case E: $j=m(\neq l)$ \\
Case B: $j\neq l$ & Case D: $j\neq m$ & Case F: $j=l(\neq m)$ \\
&  & Case G: $j\neq m$ and $j\neq l$ \\ \hline
\end{tabular}
\]

For each case we count the possible paths below and thus determine $%
U_{(i,j),(l,m)}^{3}$. Note that all paths of length three must be of
the form
\[
(i,j)\rightarrow(j,x)\rightarrow(x,l)\rightarrow(l,m),
\]
where $x\in\{i,m\}$ or $x$ is some vertex other than $i,j,l$ or $m$.

\textbf{Case A ($i=m$ and $j\neq l$)} There is the unique path that
takes place only on the vertices $\{i,j,l,m\}$:
\[
(i,j)\rightarrow(j,i)=(j,m)\rightarrow(m,l)\rightarrow(l,m),
\]
and has amplitude $\frac{2}{k}(\frac{2}{k}-1)^{2}$. The other paths
are of
the form%
\[
(i,j)\rightarrow(j,x)\rightarrow(x,l)\rightarrow(l,m),\;x\notin\{i,j,l,m\},
\]
which have amplitude $\frac{8}{k^{3}}$. If $\{j,l\}\in E(G)$ there
are $r-1$ such paths, if $\{j,l\}\notin E(G)$ there are $s-1$. Thus,
\[
U_{(i,j),(l,m)}^{3}=\frac{2}{k}\left\{ \left( \frac{2}{k}-1\right) ^{2}+%
\frac{4}{k^{2}}[s+(r-s)A_{j,l}-1]\right\} .
\]

\textbf{Case B ($i=m$ and $j=l$)} There is the unique path that
takes place only on the vertices $\{i,j,l,m\}$:
\[
(i,j)\rightarrow(j,i)\rightarrow(i,j)\rightarrow(j,i)=(l,m),
\]
which has amplitude $(\frac{2}{k}-1)^{3}$. There are $k$ other paths
of the form
\[
(i,j)\rightarrow(j,x)\rightarrow(x,j)\rightarrow(j,i)=(l,m),\;x\notin
\{i,j,l,m\},
\]
which have amplitude $\frac{8}{k^{3}}$. Thus,
\[
U_{(i,j),(l,m)}^{3}=\frac{2}{k}-1.
\]

\textbf{Case C ($i=l$ and $j\neq m$) }There are no paths which take
place only on the vertices $\{i,j,l,m\}$. There are paths
\[
(i,j)\rightarrow(j,x)\rightarrow(x,l)\rightarrow(l,m),\;x\notin\{i,j,l,m\}
\]
of amplitude $\frac{8}{k^{3}}$. If $\{j,m\}\in E(G)$ there are $r-1$
such paths, if $\{j,m\}\notin E(G)$ there are $r$. Thus,
\[
U_{(i,j),(l,m)}^{3}=\frac{4}{k^{2}}\left(
\frac{2r}{k}-A_{j,m}\right) .
\]

\textbf{Case D ($i=l$ and $j=m$) }There are no paths which take
place only on the vertices $\{i,j,l,m\}$. There are $r$ paths
\[
(i,j)\rightarrow(j,x)\rightarrow(x,l)\rightarrow(l,m),\;x\notin\{i,j,l,m\}
\]
with amplitude $\frac{8}{k^{3}}$. Thus,
\[
U_{(i,j),(l,m)}^{3}=\frac{8}{k^{3}}r.
\]

\textbf{Case E ($i\neq m,i\neq l$ and $j=m$) }If $\{i,l\}\notin
E(G)$ then there are no paths which take place only on the vertices
$\{i,j,l,m\}$. However, there are $r$ paths
\[
(i,j)\rightarrow(j,x)\rightarrow(x,l)\rightarrow(l,m),\;x\notin\{i,j,l,m\}
\]
with amplitude $\frac{8}{k^{3}}$. If $\{i,l\}\in E(G)$ there is the
unique path
\[
(i,j)\rightarrow(j,i)\rightarrow(i,l)\rightarrow(l,m)
\]
with amplitude $\frac{4}{k^{2}}(\frac{2}{k}-1)$. When $\{i,l\}\in
E(G)$ there are also $r-1$ paths
\[
(i,j)\rightarrow(j,x)\rightarrow(x,l)\rightarrow(l,m),\;x\notin\{i,j,l,m\}
\]
with amplitude $\frac{8}{k^{3}}$. Thus,%
\[
U_{(i,j),(l,m)}^{3}=\frac{4}{k^{2}}\left(
\frac{2r}{k}-A_{i,l}\right) .
\]

\textbf{Case F ($i\neq m,i\neq l$ and $j=l$) }There are two paths%
\[
(i,j)\rightarrow(j,i)\rightarrow(i,j)\rightarrow(j,m)=(l,m)
\]
and
\[
(i,j)\rightarrow(j,l)\rightarrow(l,j)\rightarrow(j,m)=(l,m)
\]
each of amplitude $\frac{2}{k}(\frac{2}{k}-1)^{2}$. There are also
$k-2$ paths
\[
(i,j)\rightarrow(j,x)\rightarrow(x,l)\rightarrow(l,m),\;x\notin\{i,j,l,m\}
\]
with amplitude $\frac{4}{k^{2}}(\frac{2}{k}-1)$. Thus,%
\[
U_{(i,j),(l,m)}^{3}=0.
\]

\textbf{Case G ($i\neq m,i\neq l,j\neq l$ and $j\neq l$)} We
consider the cases $\{j,l\}\in E(G)$ and $\{j,l\}\notin E(G)$
separately. Firstly, we consider the case $\{j,l\}\in E(G)$. There
are $r-A_{i,l}-A_{j,m}$ paths
\[
(i,j)\rightarrow(j,x)\rightarrow(x,l)\rightarrow(l,m),\;x\notin\{i,j,l,m\}
\]
of amplitude $\frac{8}{k^{3}}$. There are also $A_{i,l}+A_{j,m}$
paths of
amplitude $\frac{4}{k^{2}}(\frac{2}{k}-1)$. If they exist, these are%
\[
(i,j)\rightarrow(j,m)\rightarrow(m,l)\rightarrow(l,m)
\]
and%
\[
(i,j)\rightarrow(j,i)\rightarrow(i,l)\rightarrow(l,m).
\]
Now, if $\{j,l\}\notin E(G)$ the analysis is identical but with $s$
replacing $r$. Thus,
\[
U_{(i,j),(l,m)}^{3}=\frac{4}{k^{2}}\left\{ \frac{2}{k}\left[ s+(r-s)A_{j,l}%
\right] -(A_{i,l}+A_{j,m})\right\} .
\]

Notice that the conditions in the statement of the proposition are
mutually exclusive. Each case corresponds to $U_{(i,j),(l,m)}^{3}>0$
for one of the cases A-F above. The condition $1$ corresponds to the
case A, $2$ to C, $3$ to D, $4$ to E and $5$ to F. We have
$U_{(i,j),(l,m)}^{3}\leq0$ for all the other cases.
\end{proof}

\bigskip

We conclude with the following conjecture.

\begin{conjecture}
Let $G$ and $H$ be strongly regular graphs with the same set of
parameters. Then $G\cong H$ if and only if Sp$(S^{+}(U(G)^{3}))=$
Sp$(S^{+}(U(H)^{3}))$.
\end{conjecture}

If the conjecture is true then isomorphism of strongly regular
graphs can be tested in polynomial time.

\section{Computations}

The table below contains the parameters of the strongly regular
graphs for which we have tested the validity of the conjecture. We
have tested also the complements of these graphs. In all cases the
conjecture resulted to be
valid. The number of non-isomorphic srg$(n,k,r,s)$ is denoted by $N(n,k,r,s)$%
. The graphs srg$(16,9,4,6)$ where obtained from \cite{mc}; all
other graphs where obtained from \cite{s}:

\bigskip

\[
\begin{tabular}{||l|l||l|l||}
\hline $(n,k,r,s)$ & $N(n,k,r,s)$ & $(n,k,r,s)$ & $N(n,k,r,s)$
\\ \hline\hline
(16,6,2,2) & 2 & (35,18,9,9) & 227 \\
(16,9,4,6) & 2 & (36,14,4,6) & 180 \\
(25,12,5,6) & 15 & (36,15,6,6) & 32,548 \\
(26,10,3,4) & 10 & (40,12,2,4) & 28 \\
(28,12,6,4) & 4 & (45,12,3,3) & 78 \\
(29,14,6,7) & 41 & (64,18,2,6) & 167 \\ \hline
\end{tabular}%
\]

\bigskip

The method that we have described appears then to be successful for
strongly regular graphs. In general, there are cases in which the
method fails. We have verified that Sp$(S^{+}(U(G)^{3}))$
distinguishes all regular graphs up to $13$ vertices. However, we
have found $4$-regular graphs on $14$ vertices which are not
distinguished by our method. Two graphs for which the method fails
are drawn in the figure below:

\begin{figure}[ht]
\begin{center}
\includegraphics[scale=.35]{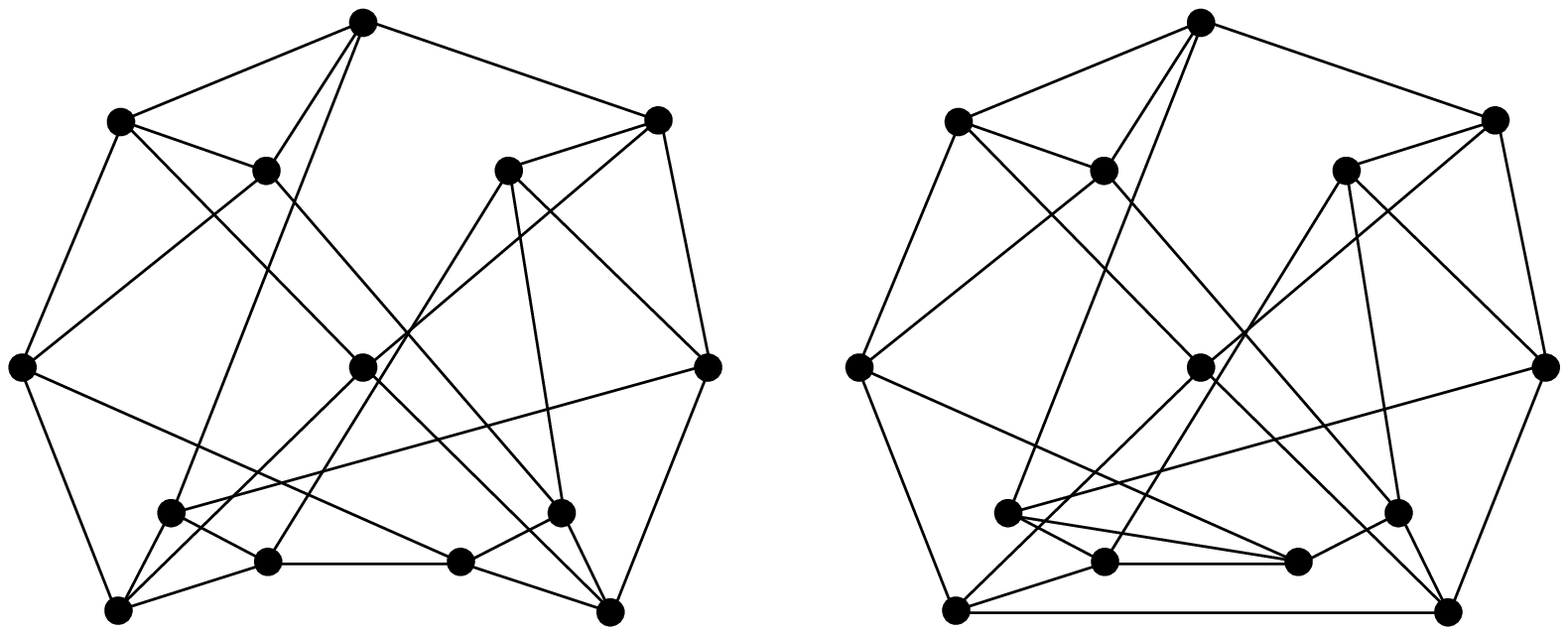}
\end{center}
\end{figure}

%

The computer programmes used for our calculations can be downloaded
from
\begin{center}
\texttt{www-users.cs.york.ac.uk/\symbol{126}wilson/qwalks.html}
\end{center}
or asked directly to the authors. These are MATLAB and C codes.

\bigskip

\noindent \textbf{Acknowledgments} We would like to thank Lalit
Jain, Chris Godsil, Toufik Mansour, Peter Cameron \ and Dennis Shasha for their
interest in this work. We would like to thank the anonymous referees
for their help in improving the content and the presentation of the
paper. One of the referees noted that there are $3$-regular graphs
on $16$ vertices that are not distinguished by the technique
described in the paper.

\end{document}